High multipole transitions in NIXS: valence and hybridization in 4$f$ systems

R.A. Gordon[1,*], G.T. Seidler[2], T.T. Fister[2], M.W. Haverkort[3], G.A. Sawatzky[4], A. Tanaka[5] and T.K. Sham[6]

[1] Dept. Of Physics, Simon Fraser University, Burnaby, BC, V5A 1S6, Canada
[2] Dept. Of Physics, University of Washington, Seattle, WA, 98105, USA
[3] II. Physikalisches institut, Universität zu Köln, Zülpicher Str. 77, D-50937 Köln, Germany
[4] Dept. Of Physics and Astronomy, University of British Columbia, Vancouver, BC, V6T 1Z1
[5] Dept. Of Quantum Matter, ADSM, Hiroshima University, Higashi-Hiroshima 739-8530, Japan
[6] Dept. Of Chemistry, University of Western Ontario, London, Ontario, N6A 5B7, Canada

Abstract - Momentum-transfer ($q$) dependent non-resonant inelastic x-ray scattering measurements were made at the $N_{4,5}$ edges for several rare earth compounds. With increasing $q$, giant dipole resonances diminish, to be replaced by strong multiplet lines at lower energy transfer. These multiplets result from two different orders of multipole scattering and are distinct for systems with simple $4f^0$ and $4f^1$ initial states. A many-body theoretical treatment of the multiplets agrees well with the experimental data on ionic La and Ce phosphate reference compounds. Comparing measurements on $CeO_2$ and $CeRh_3$ to the theory and the phosphates indicates sensitivity to hybridization as observed by a broadening of $4f^0$-related multiplet features. We expect such strong, nondipole features to be generic for NIXS from f-electron systems.

[*] Corresponding author, e-mail: ragordon@sfu.ca

The electronic, magnetic, and structural properties of rare earth materials are strongly influenced by intra- and inter-atomic interactions involving the 4f electrons. The problem of the local electronic structure of Ce provides a valuable case in point. Both $Ce^{3+}$ ($4f^1$) and $Ce^{4+}$ ($4f^0$) formally exist in ionic materials, but interaction between the 4f states and ligand or conduction states plays an important role in many valence fluctuation or heavy fermion cerium materials [1-11] such as Ce metal, $CeRh_3$, $CePd_3$, and $CeO_2$, with cerium oxides in particular continuing to attract considerable attention for their catalytic properties [12-16]. The importance of hybridization in spite of the small radial extent of the 4f wavefunctions is due to the large 4f degeneracy. Further understanding of the 4f localization behavior in this class of materials is hampered not only by the theoretical difficulties presented by, *e.g.*, competing interactions and strong hybridization, but also by the difficulty in obtaining and interpreting relevant spectroscopic information about the Ce local electronic structure.

Several core-shell spectroscopies have been applied to the 4d initial state in order to probe f-type final states in cerium compounds. At the Ce $N_{4,5}$ edge, spin-orbit splitting is small, 4f-4d interaction is large [7], and the low binding energy of the 4d states compared to 3d or 2p orbitals results in a core-hole whose effect on 4f energy levels from reduced screening of the nuclear charge is lessened. X-ray absorption spectroscopy (XAS) measurements at the $N_{4,5}$ edge however, contain both surface and bulk effects in their results and the need for a bulk-sensitive probe is understood [1]. Weak pre-edge multiplet features observed by XAS [4,6,8-10,17] at the $N_{4,5}$ edge are due to the slight mixing of localized dipole-forbidden and delocalized dipole-allowed states [17]. Reflection-mode electron energy loss spectroscopy (EELS) at low incident electron energy has also been used to probe dipole and nondipole transitions to 4f-type final states [18, 19], but such measurements are also quite surface sensitive and may have additional complications due to multiple scattering.

We revisit the phenomenon of excitations at the $N_{4,5}$ edge in cerium compounds from a new perspective: non-resonant inelastic x-ray scattering (NIXS) [20, 21]. Unlike XAS or electron spectroscopies at this edge energy, the fact that both the incident and scattered photons are relatively high energy (typically ~10 keV) ensures that NIXS is an inherently bulk-sensitive technique [22, 23]. We find that $q$-dependent NIXS provides straightforward, bulk-sensitive measurement of both dipole-allowed and dipole-forbidden $4d \rightarrow 4f$ transitions in Ce compounds. Doing so has revealed strong pre-threshold resonances that are non-dipole in character, sensitive to f-occupation, and dominate the spectrum at high-$q$.

The double-differential cross-section in a NIXS measurement [24, 25] is:

$$\frac{d^2\sigma}{d\Omega d\omega} = \left(\frac{d\sigma}{d\Omega}\right)_{Th} S(\vec{q},\omega), \qquad (1)$$

where $(d\sigma/d\Omega)_{Th}$ is the Thomson scattering cross section and $S(\mathbf{q}, \omega)$ is the dynamical structure factor. Note that

$$S(\vec{q},\omega) = \sum_f \left| \langle f | e^{i\vec{q}\cdot\vec{r}} | i \rangle \right|^2 \delta(E_i - E_f + \hbar\omega) \qquad (2)$$

where $\mathbf{q}$ is the momentum transfer, $\mathbf{k}_f - \mathbf{k}_i$, and $\hbar\omega$ is the photon energy loss. When $q$ ($|\mathbf{q}|$) is sufficiently small, the matrix element reduces to a dipole transition as is measured in XAS. However, even for low energy loss, NIXS is a bulk-sensitive technique: the penetration of the sample is determined by the incident photon energy. Increasing $q$ results in new contributions to $S(q,\omega)$ from non-dipole transitions, allowing characterization of final states inaccessible to XAS [25-29]. The development of third-generation synchrotron sources and multi-element crystal analyzer spectrometers has resulted in a rapid increase in applications of valence and core-shell NIXS [20, 30-36]. We have used such to examine the NIXS contribution from the 4$d$ initial states of isotypic compounds LaPO$_4$ and CePO$_4$ [37] as representatives of 4$f^0$ and 4$f^1$ occupation for theoretical treatment and compare the results to data on the compounds CeO$_2$ and CeRh$_3$ where the 4$f$ states are considered to be interacting with ligand or conduction states.

Commercial powders (Alfa) of LaPO$_4$, CePO$_4$ and CeO$_2$ were pressed into pellets approximately 0.25 mm thick and sectioned into pieces for mounting. The CeRh$_3$ sample was prepared from 99.9% pure elements (Alfa) by arc-melting and subsequent anneal under vacuum for 1 week at 800ºC. The final mass of the arc-melted bead was within 1.5% of the stoichiometric value. The bead was cracked under liquid nitrogen to liberate a pale-metallic-pink rectangular fragment roughly 2.5 x 1.5 x 0.75 mm$^3$ from the interior of the bead.

Experiments were carried out using the LERIX end-station [38] at the PNC/XOR insertion device beamline, Sector 20 at the Advanced Photon Source, Argonne National Laboratory. For these measurements, toroidal focusing of the x-ray beam from the Si(111) double-crystal monochromator was performed, giving approximately 10$^{13}$ photons/s at ~10 keV incident on the samples. The LERIX instrument was configured to have a full 19-detector (spherically-bent Si(555)) array measuring, independently, momentum transfers ranging from 0.8 to 10.1 Å$^{-1}$. Penetration depths of the x-rays varied from 6 μm (CeRh$_3$) to 14 μm (phosphates), ensuring truly bulk-sensitive measurements. The measurements had a resolution of 1.3 eV. Samples were oriented so that the x-rays were incident approximately 6° to the surface of the pellet or metal section. For CePO$_4$ and CeO$_2$, the 0.8 Å$^{-1}$ detector had a signal level that was greatly reduced compared to the next lowest $q$ - likely to due surface roughness of the pressed-powder samples when measured near grazing angle - and was neglected.

The $N_{4,5}$ contribution to the NIXS intensities of the phosphates are shown in fig. 1. At low-$q$, the spectra are dominated by the giant dipole resonance as observed for similar compounds by XAS and EELS techniques [e.g. 4, 6, 8, 18]. Due to low counting statistics (2 hours total scan time per sample per edge), the weak pre-edge multiplet structures observed by XAS or EELS in this dipole regime are not resolved here. With increasing $q$, the dipole resonance diminishes and the growth of a new multiplet structure manifests as higher-order (i.e. beyond dipole) transitions begin to contribute. Note that the multiplet features arrive with two different q-dependencies, indicating the presence of two different non-dipole scattering channels. This is well-illustrated by the La compound. At 3.9 Å$^{-1}$, two peaks have emerged 13 eV and 16 eV below the giant resonance. At 7.7 Å$^{-1}$, additional features begin to appear ~15 eV and 18 eV below the giant resonance. By $q \sim 8$ Å$^{-1}$, the contribution to the spectrum from the dipole resonance is heavily reduced [25, 39] and the NIXS spectra near the 4$d$ binding energy are dominated by the pre-threshold structure. These results show a strong sensitivity of the non-dipole multiplet structure to the 4$f$-occupation. As the final states for these transitions are apparently localized (given their discrete nature), we infer that the non-dipole transitions are atomic-like $4d^{10}4f^n \rightarrow 4d^94f^{n+1}$ of differing symmetry (e.g. $^1S \rightarrow{}^1F$, octupole; $^1S \rightarrow{}^1H$, triakontadipole; etc) and that the different multiplet spectra represent the difference between $4f^0$ (La$^{3+}$) and $4f^1$ (Ce$^{3+}$) initial states.

Calculations of the NIXS spectra at the Ce $N_{4,5}$ edge were performed for the two ionic configurations Ce$^{3+}$ (initial state $4d^{10}4f^1$) and Ce$^{4+}$ ($4d^{10}4f^0$). The method of calculation was a local many-body approach similar to that used previously to understand the $q$-dependence of $d$-$d$ excitations in NiO and CoO [39]. Hartree-Fock (HF)-based radial wavefunctions were used in evaluating the expression for the dynamical structure factor, S($\mathbf{q}$, ω) in eq. (2). Expanding the exponential term in terms of spherical harmonics yields

$$e^{i\vec{q}\cdot\vec{r}} = \sum_{k=0}^{\infty}\sum_{m=-k}^{k} i^k (2k+1) j_k(qr) C_m^{(k)*}(\theta_q,\phi_q) C_m^{(k)}(\theta_r,\phi_r) \quad (3)$$

with the $k^{th}$-order spherical Bessel function, $j_k(qr)$, for the $q$-dependent radial component and the angular dependence, $C_m^{(k)} = \sqrt{4\pi/2k+1} \times Y_{km}$, averaging out since the samples are powder or polycrystalline. Inserting eq. (3) into eq. (2) yields

$$S(q,\omega) = \sum_f \sum_{k=0}^{\infty} D_k |<f(r)|j_k(qr)|i(r)>|^2 \delta(E_i - E_f + \hbar\omega) \quad (4)$$

with the HF radial wavefunctions $f(r)$, $i(r)$, and coefficients, $D_k$. As the f-wavefunction is odd under inversion and the d-wavefunction even, only transitions of odd parity and f-d=1 ≤ k ≤ d+f = 5 contribute to the intensity (f and d being the orbital angular momentum quantum numbers for f and d orbitals). In eq. (4) then, only terms k = 1, 3 and 5 survive, with k = 1 being the dipole transition. These terms for $Ce^{4+}$ and $Ce^{3+}$ are displayed in fig. 2 as are the calculated $S(q,\omega)$.

The strength of the different k-term contributions to $S(q,\omega)$ is dependent on $q$ as per eq. (4) [25, 38]. Agreement is good between the calculations and data, but not exact. In general, one finds that the change from dipole (k=1) to octupole (k=3) or higher transitions in the calculation occurs at higher $q$ values than in the experiment. This is related to the general finding that within a HF calculation, the radial wave functions are less-extended than in real systems, which results in the calculated Slater integrals being too large. It is well established [40, 41] that the Slater integrals for a real system involving 4$f$ orbitals and other core orbitals can be quite strongly reduced compared to those obtained from HF calculations. One reason for this is strong correlation effects between the electrons in spatially overlapping 4$d$ and 4$f$ wave functions and the very strong Coulomb and exchange interactions. The very large multiplet energy spread for the final states in configurations like $4d^9 4f^1$ also can lead to a partial breakdown of the usefulness in describing the full multiplet structure in terms of a single set of Slater integrals. These issues can be studied with configuration interaction approaches well known in atomic physics [42]. It will be interesting to see if such approaches can at least partly resolve the difference in the $q$-dependent cross-over from a dipole to a multipole transition-dominated NIXS spectrum. This requires however a detailed and systematic study of numerous excited state configurations which could contribute and will be reported on in future work.

The dipole features above 120 eV do not agree well with experiment. The calculated spectra are sharper and show less structure for this regime. This problem has been realized before for the description of XAS [43]. The $4d$ core-hole/$4f$ valence Coulomb interaction results in an extensive multiplet structure where some states are pushed to quite high energy. These high energy final states interact strongly with continuum states that are equal or lower in energy than the highest multiplets of the $4d^9 4f^{n+1}$ configuration. This interaction results in the characteristic shape of the observed higher multiplets (giant dipole resonance). For $Ce^{4+}$ (or $La^{3+}$), which has a singlet S initial state, the final state due to a dipole (k=1) excitation will have P character whereas an octupole excitation (k=3) will excite into states with F character. The state with P character is the highest multiplet of the $4d^9 4f^1$ configuration. One can see in fig. 1 that the energy difference between these different multiplets is of the order of 15 eV.

The fact that the lower-energy, non-dipole resonances are not broadened by interaction with the continuum simplifies both their detection by NIXS and their explanation. While the preceding calculations have been specific to the case of $Ce^{3+}/Ce^{4+}$, it is important to recognize that the physical issues involved are quite general. We anticipate that analogous multiplet-like features will be observed in any f-electron system, and that the issues of valence, hybridization, and final state effects will each play a role in the resulting high-$q$ NIXS spectra.

We have limited the present theoretical treatment to the ionic compounds but must still consider the influence of 4$f$ hybridization. Figure 3 contains the $q$-dependent NIXS data for $CeO_2$. As with the phosphates, there is a change with increasing $q$ from a giant dipole resonance to a multiplet structure at lower energy. Overall, the data for $CeO_2$ resembles that for a 4$f^0$ system. The low-$q$ giant resonance resembles that observed by XAS [9, 10]. Near 4.6 Å$^{-1}$, two peaks appear ~ 15 eV below the giant resonance and additional features appear near 7.7 Å$^{-1}$. However, compared to $LaPO_4$, the features are broadened. This is better illustrated in fig. 4, where the high-$q$ data for $LaPO_4$, $CePO_4$, $CeO_2$ and $CeRh_3$ are compared directly. Multiplet features can be resolved in $CeO_2$, but do appear broader than the (shifted) La compound. This suggests that the 4$f$ states are largely unoccupied in $CeO_2$ but with some hybridization (between Ce 4$f$ and O 2$p$ states [15]). This is consistent with electronic structure calculations [15], and with oxygen K-edge data on $CeO_2$ and $Ce_2O_3$ [16], where a pre-edge peak associated with 4$f^0$ character is readily distinguished from a pre-edge feature associated with 4$f^1$ interacting with O 2$p$ states. Cerium 4$f$ states in intermetallic $CeRh_3$ are considered to be more strongly hybridized than cerium in $CeO_2$ (becoming band-like [44]). This is observable in the NIXS $N_{4,5}$ region by the lack of resolvable features in a pre-threshold peak that has a maximum at an energy loss comparable to, and is asymmetric to lower energy loss more like the 4$f^0$ materials than the 4$f^1$ cerium phosphate. It is evident, then, that the manifestation of hybridization effects (covalency) in f-electron compounds is readily observed by broadening in pre-threshold multiplet features. It is in this area – covalence effects – that future theoretical treatments on $N_{4,5}$ NIXS must be directed in order to better understand 4$f$ interactions in these materials.

Before concluding, it is interesting to consider the possible future application of $N_{4,5}$

NIXS to single crystals of *f*-electron materials. For single crystal measurements, the angular dependence will remain in eq. (3) and S($q$,ω) will provide additional information about the local electronic properties. XAS and core shell NIXS measurements on single crystals of low-Z compounds show a strong sensitivity to anisotropy in the local electronic structure, such as those resulting from the details of chemical bonding [45-47]. The sensitivity of $N_{4,5}$ NIXS to such effects would provide a measure of the involvement of *f*-electrons in bonding. The evolution of the *l*- and *m*- dependent selection rules at high *q* also suggest possible sensitivity of $N_{4,5}$ NIXS to the ground-state symmetry, or orbital or magnetic ordering. Single crystal effects in $N_{4,5}$ NIXS may be modeled by methods similar to those used recently [39] to treat NIXS from weakly bound electrons in single crystal NiO and CoO [20].

In summary, we have examined a number of cerium-containing compounds near the Ce $N_{4,5}$ resonance using momentum-dependent non-resonant inelastic x-ray scattering. The spectra demonstrate a transition with increasing momentum transfer from dipole-like behavior similar to XAS measurements to growth of momentum-dependent multiplet features. The resulting multiplet features are valence and hybridization-sensitive. Results for a $4f^1$ ionic compound, $CePO_4$, are readily distinguishable from a $4f^0$ ionic compound, $LaPO_4$. Hartree-Fock-based calculations are able to model the multiplet structures for the ionic compounds with good agreement of the *q*-dependence. Broadening of the multiplet features is signature behavior of interaction between the f –states and conduction or ligand states, as evidenced by the behavior of two classic 4*f*-interacting materials, $CeO_2$ and $CeRh_3$. The ability to observe these *q*-dependent pre-threshold resonances outside the dipole limit is expected to be generic and thus provides a new spectroscopic tool for investigations of the physics and chemistry of f-electron systems.

This work was supported by the U.S. Dept. of Energy - Basic Energy Sciences, the Office

of Naval Research, the Bosack and Kruger Foundation, the Natural Sciences and Engineering Research Council of Canada, the Canadian Institute for Advanced Research and the Deutsche Forschungsgemeinschaft. We would like to thank J. Garrett of BIMS, McMaster University for the $CeRh_3$ sample and K. Nagle of the University of Washington for answering numerous LERIX-related questions. PNC/XOR facilities at the Advanced Photon Source, and research at these facilities, are supported by the U.S. Dept. of Energy - Basic Energy Sciences, a major facilities access grant from NSERC, the University of Washington, Simon Fraser University and the Advanced Photon Source.
REFERENCES

[1] PARLEBAS J.C. and KOTANI A., *J. Electr. Spectr. Rel. Phenom.*, **136** (2004) 3.

[2] KANAI K and SHIN S., *J. Electr. Spectr. Rel. Phenom.*, **117 – 118** (2001) 383.

[3] VAVASSORI P. *et al.*, *Phys. Rev. B*, **52** (1995) 16503.

[4] PETERMAN D.J., WEAVER J.H. and CROFT M., *Phys. Rev. B*, **25** (1982) 5530.

[5] NAKAZAWA M. and KOTANI A., *J. Phys. Soc. Japan*, **71** (2002) 2804.

[6] KALKOWSKI G. *et al.*, *Phys. Rev. B*, **32** (1985) 2717.

[7] JO T. and KOTANI A., *Phys. Rev. B*, **38** (1988) 830.

[8] KOTANI A. *et al.*, *Phys. Rev. B*, **40** (1989) 65.

[9] MATSUMOTO M. *et al. Phys. Rev. B*, **50** (1994) 11340.

[10] MATSUMOTO M. *et al.*, *J. Electr. Spect. Rel. Phenom.*, **78** (1996) 179.

[11] SHAM T.K., GORDON R.A. and HEALD S.M., *Phys. Rev. B*, **72** (2005) 035113.

[12] FONDO E. *et al.*, *J. Synchrotron Rad.*, **6** (1999) 34.

[13] PETIT L. *et al.*, *Phys. Rev. B*, **72** (2005) 205118.

[14] LOSCHEN C. *et al.*, *Phys. Rev. B*, **75** (2007) 035115.

[15] SKORODUMOVA N.V. *et al. Phys. Rev. B*, **64** (2001) 115108.

[16] MULLINS D.R., OVERBURY S.H. and HUNTLEY D.R., *Surf. Sci.*, **409** (1998) 307.

[17] JO T., *Prog. Theor. Phys.*, **101** (1990) 303.

[18] NETZER F.P., STRASSER G. and MATTHEW J.A.D., *Phys. Rev. Lett.*, **51** (1983) 211.

[19] OGASAWARA H. and KOTANI A., *J. Electr. Spectr. Rel. Phenom.*, **78** (1996) 119.

[20] LARSON B.C. *et al.*, *Phys. Rev. Lett.*, **99** (2007) 026401.

[21] TOHJI K. and UDAGAWA Y., *Phys. Rev. B*, **36** (1987) 9410.

[22] KRISCH M. and SETTE F., *Surf. Rev and Lett.*, **9** (2002) 969.

[23] BERGMANN U., GLATZEL P. and CRAMER S.P., *Microchem. J.*, **71** (2002) 221.

[24] SCHULKE W., *Handbook of Synchrotron Radiation*, edited by G.S. Brown and D.E. Moncton (North-Holland, Amsterdam, 1991), Vol. 3.

[25] SOININEN J.A., ANKUDINOV A.L. and REHR J.J., *Phys. Rev. B*, **72** (2005) 045136.

[26] STERNEMANN C. *et al.*, *Phys. Rev. B*, **68** (2003) 035111.

[27] KRISCH M., SETTE F., MASCIOVECCHIO C. and VERBENI R., *Phys. Rev. Lett.*, **78** (1997) 2843.

[28] FENG Y.J. *et al.*, *Phys. Rev. B*, **69** (2004) 125402.

[29] HAMALAINEN K. *et al.*, *Phys. Rev. B*, **65** (2002) 155111.

[30] FISTER T.T. *et al.*, *Phys. Rev. B*, **74** (2006) 214117.

[31] STERNEMANN H. *et al.*, *Phys Rev B*, **75** (2007) 245102.

[32] BALASUBRAMANIAN M. *et al.*, *App. Phys. Lett.*, **91** (2007) 031904.

[33] LEE S.K. *et al.*, *Phys. Rev. Lett.*, **98** (2007) 105502.

[34] MAO W.L. *et al.*, *Science*, **302** (2003) 425.

[35] SOININEN J.A. *et al.*, *J. Phys.-Cond. Matter*, **18** (2006) 7327.

[36] WERNET P. *et al.*, *Science*, **304** (2004) 995.

[37] NI Y., HUGHES J.M. and MARIANO A.N., *Am. Mineral.*, **80** (1995) 21.

[38] FISTER T.T. *et al.*, *Rev. Sci. Instrum.*, **77** (2006) 063901.

[39] HAVERKORT M.W. *et al.*, arXiv:0705.4637 (2007).

[40] DAGYS R. and KANCEREVIČIUS A., *Phys. Rev. B,* **53** (1996) 977.

[41] SUGAR J., *Phys. Rev. A*, **6** (1972) 1764.

[42] COWAN R.D., *The Theory of Atomic Structure and Spectra*, (University of California Press, Berkeley) 1981.

[43] DEHMER J.L. *et al.*, *Phys. Rev. Lett.*, **26** (1971) 1521.

[44] WESCHKE E. *et al.*, *Phys. Rev. Lett.*, **69** (1992) 1792.

[45] STOHR J., *NEXAFS Spectroscopy*, Springer Series in Surface Sciences, Vol. 25 (Springer-Verlag, New York) 1996.

[46] WATANABE N. *et al., Appl. Phys. Lett.,* **69** (1996) 1370.

[47] MATTILA A., *et al., Phys. Rev. Lett.,* **94** (2005) 247003.

Figure Captions

FIG 1. The momentum ($q$) dependence of the contribution to $S(q,\omega)$ from the 4$d$ initial state for a powder sample of a) LaPO$_4$ and b) CePO$_4$ as representatives of ionic 4$f^0$ and 4$f^1$ compounds, respectively.

FIG 2. Calculated momentum-dependent multiplet structures for a) Ce$^{4+}$ (4$f^0$) and b) Ce$^{3+}$ (4$f^1$) The more intense components of the dipole terms in the calculated $S(q,\omega)$ have been truncated for clarity. The individual multipole terms k = 1 (dipole, dashed), k = 3 (solid) and k = 5 (dash-dot) contributing to the calculated $S(q,\omega)$ at the N$_{4,5}$ edge are shown in the upper panel.

FIG 3. The momentum ($q$) dependence of the contribution to $S(q,\omega)$ from the 4$d$ initial state for a powder sample of CeO$_2$.

FIG 4. A comparison of the high-$q$ average multiplet structures at the N$_{4,5}$ region of the measured NIXS spectra for LaPO$_4$, CeO$_2$, CeRh$_3$, and CePO$_4$ showing the influence of 4$f$ hybridization on CeO$_2$ and CeRh$_3$.

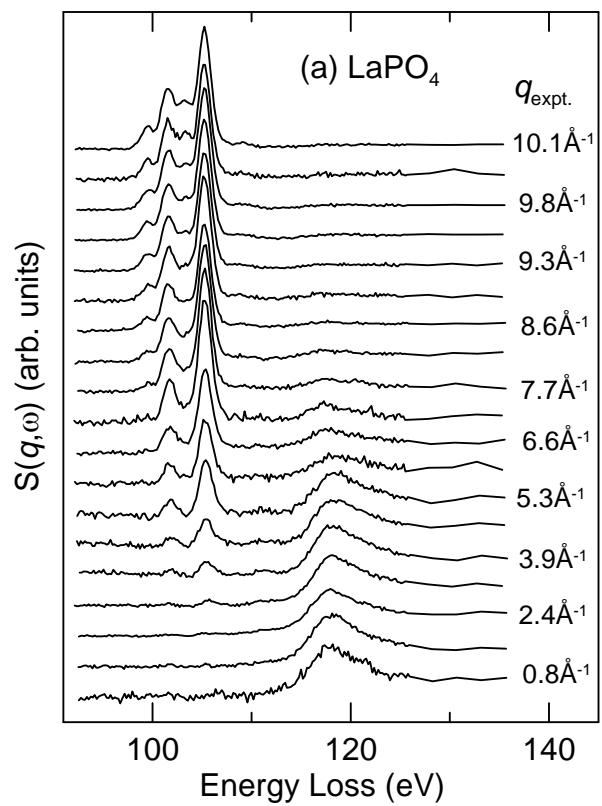

Fig 1a

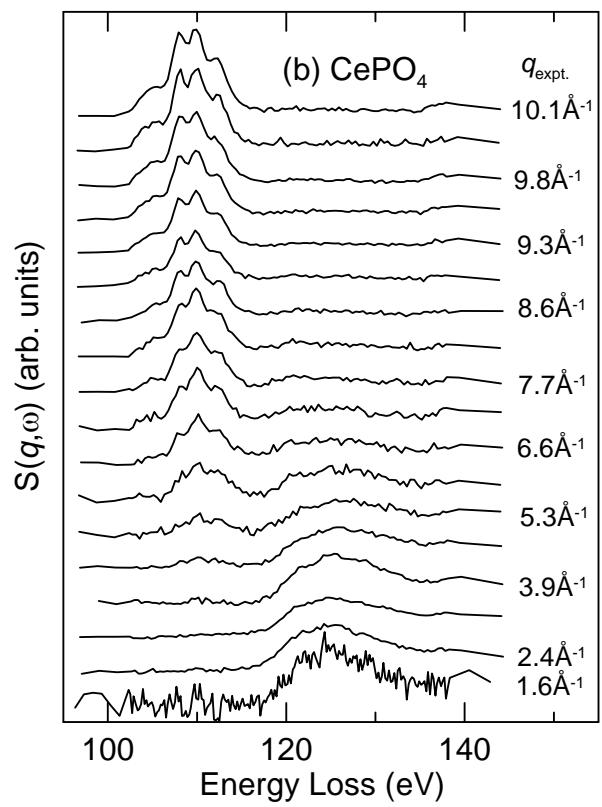

Fig. 1b

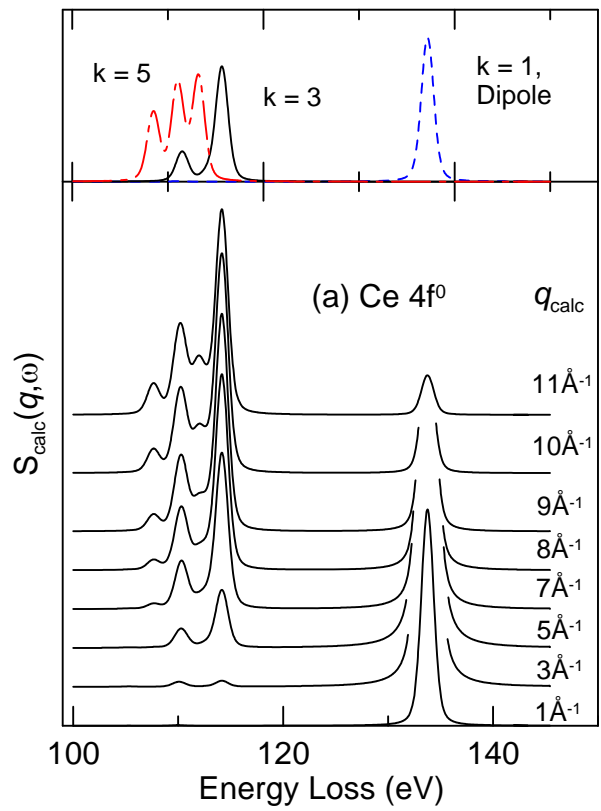

Fig 2a

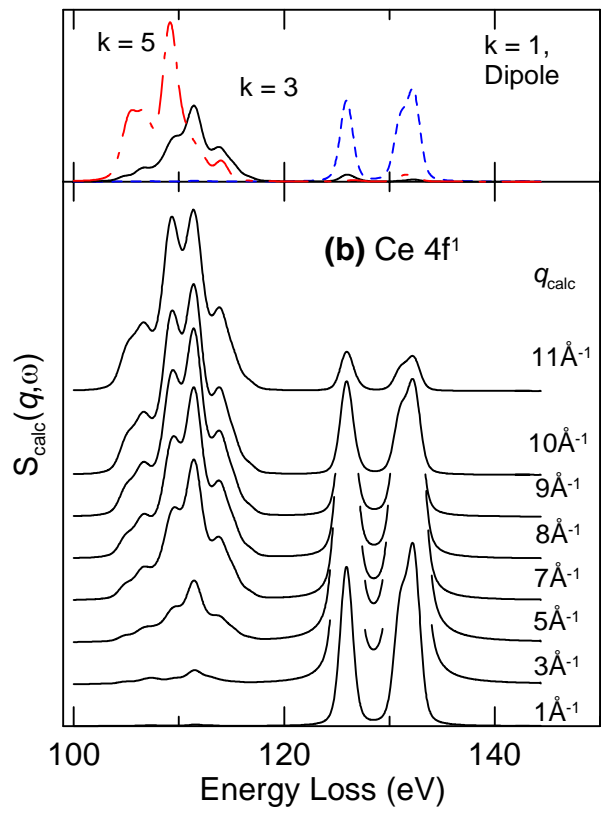

Fig 2b

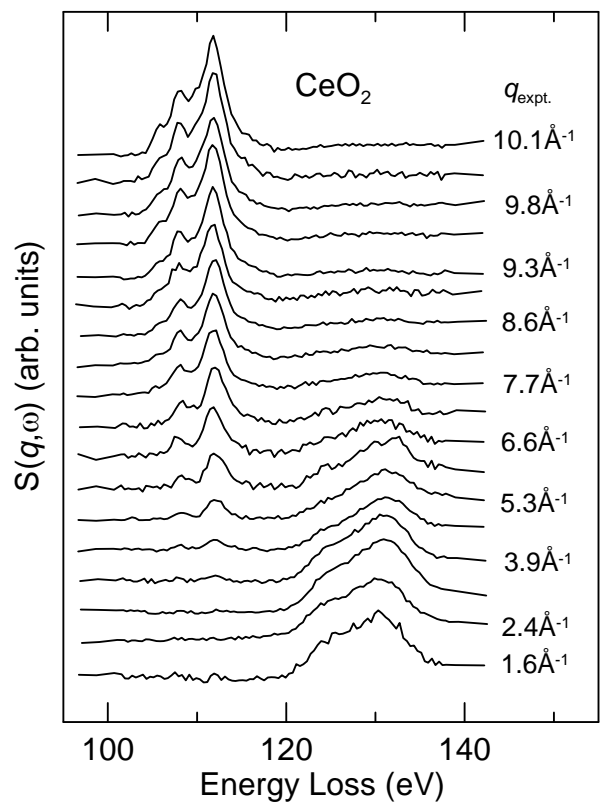

Fig 3

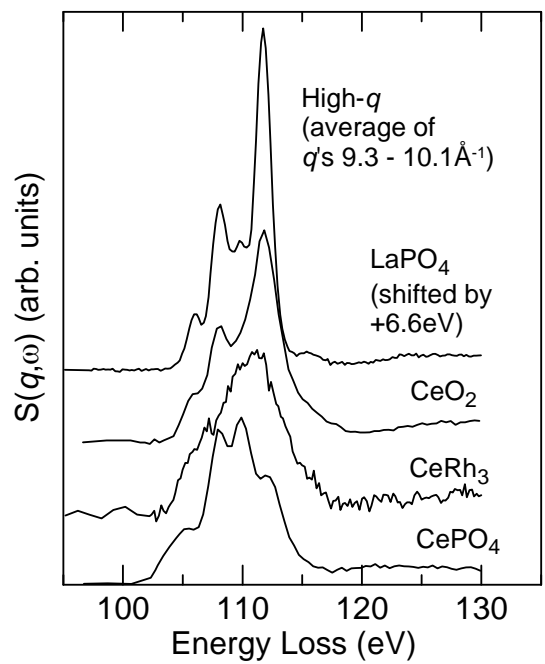

Fig 4